\documentclass[journal]{IEEEtran}

\usepackage{myStyleIEEE}
\allowdisplaybreaks
\usepackage{mathabx}
\IEEEoverridecommandlockouts

\begin{document}

\title{Improving Stability of Low-Inertia Systems using Virtual Induction Machine Synchronization for Grid-Following Converters}

\renewcommand{\theenumi}{\alph{enumi}}

\author{Ognjen~Stanojev,~\IEEEmembership{Student~Member,~IEEE,}
        Uros~Markovic,~\IEEEmembership{Member,~IEEE,} \\
        Petros~Aristidou,~\IEEEmembership{Member,~IEEE,}
        Gabriela~Hug,~\IEEEmembership{Senior~Member,~IEEE,}\vspace{-0.35cm}% <-this % stops a space\vspace{-0.35cm}
\thanks{Research supported by NCCR Automation, a National Centre of Competence in Research, funded by the Swiss National Science Foundation (grant number 180545).}
}

\maketitle
\IEEEpeerreviewmaketitle

%ABSTRACT
\begin{abstract}
This paper presents a novel strategy for the synchronization of grid-following Voltage Source Converters (VSCs) in power systems with low rotational inertia. The proposed synchronization unit is based on emulating the physical properties of an induction machine and capitalizes on its inherent grid-friendly properties such as self-synchronization, oscillation damping, and standalone capabilities. To this end, the mathematical model of an induction machine is analyzed and reformulated to obtain the unknown grid frequency by processing the voltage and current measurements at the converter output. This eliminates the need for a Phase-Locked Loop (PLL) unit, traditionally employed in grid-following VSC control schemes, while simultaneously preserving the system- and device-level control. Furthermore, we provide the appropriate steps for obtaining an index-1 DAE representation of the induction-machine-based synchronization unit, suitable for stability analysis. Our analysis shows that replacing the PLL unit with the virtual induction machine-based synchronization may considerably improve the stability of systems with grid-following converters and facilitate the frequency containment. Furthermore, we validate the performance of the proposed synchronization unit through simulations and provide recommendations for its tuning.
%The EMT simulations validate the mathematical principles of the proposed model, whereas a small-signal analysis provides guidelines for appropriate control tuning and reveals interesting properties pertaining to the nature of the underlying operation mode.
\end{abstract}

%INDEX TERMS

\section{Introduction}
\lettrine[lines=2]{E}{fforts} to reduce the environmental impact of the energy sector are bringing about increasing amounts of renewable energy sources to power systems, which are typically interfaced to the grid via Voltage Source Converters (VSCs)~\cite{Milano2018}. In contrast to Synchronous Generators (SGs) which synchronize to the grid through the electromagnetic coupling between the rotor field and the grid-imposed stator currents\cite{Kundur}, interfacing and synchronization characteristics of the converter-based resources are determined solely by the implemented control logic~\cite{SyncReview2020}. Control strategies for VSCs are commonly categorized into two groups\cite{Poolla2019,li2022revisiting}: (i) grid-forming controls that establish a voltage vector (frequency and voltage magnitude) at the interface and (ii) grid-following control mode that assumes a constant voltage at the interface irrespective of the converter power injection. Synchronization of grid-following inverters, which due to their simple structure have already been widely used for integrating wind and solar energy, is the main focus of this work.

In order to regulate a grid-connected VSC in the grid-following mode, a control structure comprising system-level control, with outer control loops and a synchronization unit, and device-level control, with inner control loops, has become prevalent in the industry for providing adequate VSC voltage, active and reactive power outputs \cite{Malik2017,UrosStability2021}. Additionally, the standard of having a Phase-Locked Loop (PLL) as the synchronization unit has early been established \cite{Chung2000}, together with its numerous variants \cite{Karimi2004,Golestan2014NewPLL,Golestan2018,Dongsheng2020}. However, despite being widely utilized for frequency estimation, this additional, inherently nonlinear estimator introduces additional complexity to the system. As its input signal undergoes fast electromagnetic transients, the PLL can experience numerical issues and be affected by jumps and discontinuities following discrete events in the system such as faults or line outages \cite{Junru2022}. Moreover, by the nature of its design, it introduces a non-negligible delay which can limit the performance of controls depending on frequency estimation~\cite{SyncReview2020}, and in addition, may be extremely difficult to tune \cite{Bo2016}. Several publications have recognized the impact of PLLs on the operation of non-synchronous generation \cite{Luna2015, Hu2017}, but also highlighted the potential instabilities arising from high penetration of power converters employing such synchronization devices \cite{Goksu2014,Bizzarri2017,Qi2019,UrosStability2021}.

Recent studies have addressed some of the aforementioned issues by developing PLL-less converter regulation in the form of power-synchronization \cite{LZhang2010}, virtual oscillator control \cite{JDH+14,SS+19} and self-synchronizing \textit{synchronverters} \cite{Zhong2014}. Although the proposed methods demonstrate numerous advantages, drawbacks have also been observed. Namely, the power-synchronization is mostly focused on VSC-HVDC applications and faces challenges with weak AC system connections. Virtual oscillator control faces obstacles in terms of reference tracking, whereas the synchronverter concept still requires a back-up PLL and improvements in operation under unbalanced and distorted grid voltages. Furthermore, all aforementioned controls apply to the grid-forming operation mode. While grid-forming VSCs are an integral part of a future low-inertia power grid~\cite{Lasseter2020}, the existing systems are primarily composed of renewable generation interfaced to the network via grid-following inverters. Moreover, grid-forming converters may demonstrate side-band oscillations in stiff and series compensated grids~\cite{Gaoxiang2020,Shike2020}.

Contrarily, a recently proposed VSC control method under the name of \textit{inducverter} introduces the notion of a grid-connected converter operating under Induction Machine (IM) working principles and without a dedicated PLL unit. It was first introduced in \cite{Ashabani2016}, and later further analyzed in \cite{VIG2020,UrosIREP}. Despite the concept still being at its early stages, it can potentially resolve the issues associated with the conventional PLL-based synchronization loop. However, \cite{Ashabani2016} proposes a control design that integrates the system-level controller and synchronization unit into one compact structure. As a consequence, the frequency regulation and stabilization properties are attributed to the inducverter, whereas these functionalities are primarily a consequence of the implemented PI-based droop power control. Additionally, the controller is implemented in a hybrid $abc$-$dq$ frame, where the $dq$-axis current references are obtained according to the real and reactive power errors and translated to $abc$ voltage references through an adaptive virtual impedance in the $abc$ frame. This significantly complicates the analytical representation of the model and its analysis. As a consequence, for the controller tuning it is suggested to revert back to adopting the parameters of an existing induction machine. Finally, the fact that the cascaded inner control loop is replaced by a simple adaptive lead or lag compensator raises concerns in terms of fast voltage reference tracking and overcurrents during transients.

In this work, we build upon the Virtual Induction Machine (VIM) concept, analyzed in \cite{Ashabani2016,UrosIREP,VIG2020}, to design a novel synchronization unit for grid-following VSCs with improved stability and frequency response characteristics compared to the commonly employed PLL unit. More precisely, the contributions of this paper can be summarized as follows: 
\begin{itemize}
    \item In contrast to \cite{Ashabani2016,VIG2020} where the VSC control design has been entirely reformulated to accommodate the IM model, we show that the VIM concept can be incorporated into the grid-following converter control as an independent synchronization unit, with system and device-level control preserved. Therefore, potential integration of this concept into the existing grid-following VSCs is substantially facilitated. 
    \item This paper reformulates the mathematical principles of an emulated induction machine from \cite{UrosIREP} and extends them by deriving an appropriate index-1 DAE representation of such grid-connected VSC control scheme. This allows for a detailed small-signal analysis, which aids control tuning and understanding of the potential benefits of the VIM synchronization.
    \item While the previous works focus on individual converter operation modes in a single-machine infinite-bus setting, our study considers transient interactions between SGs and grid-forming and grid following VSCs. Furthermore, using the developed small-signal model, the stability of large-scale low-inertia power systems with deployed VIM-based grid-following units is analyzed.
\end{itemize}
 The results reveal that replacing the PLL by a VIM in grid-following VSCs results in improved frequency response and stability properties, resembling the characteristics of the grid-forming controls. Furthermore, time-domain simulations demonstrate that the VIM-based grid-following converters retain accurate active and reactive power tracking capabilities, are resilient to short circuits, and possess the ability to operate even after being disconnected from the main grid.

The remainder of the paper is structured as follows. In Section~\ref{sec:vim_strategy}, the mathematical model of an induction machine and the VIM synchronization principles are presented. Section~\ref{sec:modeling} describes the implementation of the VIM-based synchronization unit into a state-of-the-art VSC controller, as well as the model formulation as an index-1 DAE system. Section~\ref{3.1sec:sim_results} showcases the EMT simulation results and model validation, whereas Section~\ref{3.1sec:stab_analysis} provides an insightful small-signal stability analysis. Finally, Section~\ref{sec:5} discusses the outlook of the study and concludes the paper.

\section{Virtual Induction Machine Strategy} \label{sec:vim_strategy}

\subsection{Induction vs. Synchronous Machine: Operating Principles} \label{subsec:IMSM}
The physical mechanism behind the machine rotor movement and the subsequent synchronization to the grid are the most notable differences between the synchronous and induction machine. While the SM always operates at synchronous speed, the IM relies on a mismatch between the synchronous speed $\omega_s\in\R_{\geq0}$ and the machine's rotor speed $\omega_r\in\R_{\geq0}$ to operate, i.e., the slip $\nu\in\R_{>0}$:
\begin{equation}
    \nu \coloneqq \frac{\omega_s - \omega_r}{\omega_s} = \frac{\omega_\nu}{\omega_s},
\end{equation}
with $\omega_\nu\coloneqq \omega_s - \omega_r$ denoting the slip frequency. Furthermore, contrary to synchronous generators, induction machines do not have an excitation system in the rotor. Thus, the ElectroMagnetic Force (EMF) induced in the rotor of an IM is a consequence of its rotation and the subsequent change of the magnetic flux linkage through the circuit. Given that the rotor is closed through an external resistance or a short-circuit ring, the induced EMF generates a current flow in the rotor conductor, which finally produces the synchronizing torque that drives the movement of the rotor. Hence, the IM can never reach the synchronous speed, since there would be no EMF in the rotor to continue its movement. Neverthless, the IMs are typically designed to operate with small slip values and therefore, the discrepancy between the synchronous speed and the IM rotation speed is negligible.

Considering the previously described properties, it can be observed that the IM with an arbitrary initial rotor speed close to the synchronous speed has a self-synchronizing capability, i.e., has the potential to synchronize with a grid of an unknown frequency and voltage magnitude. This implies that a VIM-based synchronization unit has the potential to replace the traditional PLL in the converter control design and eliminate its inherent drawbacks pertaining to time delay and stability margins. Experimental and hardware-in-the-loop tests conducted in \cite{Ashabani2016} provide evidence that confirms the validity of the approach and proves the practical viability of this technology. Nonetheless, such implementation should not be confused with system-level regulation, e.g., droop control, virtual synchronous machine or virtual oscillator control, as it does not inherently yield an emulation of inertia or frequency oscillation damping. Such services could easily be provided by an appropriate outer control loop, as will be shown later. Furthermore, while an induction machine does not have reactive power control capabilities, this functionality can still be embedded into the VIM-synchronized VSCs using an appropriate system-level control design.

\subsection{Induction Machine Emulation Strategy} \label{subsec:IME}

For the purpose of emulating the operating principles of an IM through VSC control, let us observe a dynamical model of an IM \cite{Kundur} in a synchronously-rotating $dq$-frame and SI units:
\begin{subequations}
\label{eq:Vdsr}
\begin{align}
    v_s^d &= R_s i_s^d + \dot{\psi}_s^d - \omega_s \psi_s^q, \label{eq:Vds} \\ 
    v_s^q &= R_s i_s^q + \dot{\psi}_s^q + \omega_s \psi_s^d,  \label{eq:Vqs} \\ 
    v_r^d &= 0 = R_r i_r^d + \dot{\psi}_r^d - \omega_\nu \psi_r^q,  \label{eq:Vdr} \\ 
    v_r^q &= 0 = R_r i_r^q + \dot{\psi}_r^q + \omega_\nu \psi_r^d,  \label{eq:Vqr}
\end{align}
\end{subequations}
where $v_s=\left(v_s^d,v_s^q\right)\in\R^2$, $v_r=\left(v_r^d,v_r^q\right)\in\R^2$, $\psi_s=\left(\psi_s^d,\psi_s^q\right)\in\R^2$ and $\psi_r=\left(\psi_r^d,\psi_r^q\right)\in\R^2$ are the vectors of stator and rotor voltages and flux linkages, respectively, and $R_s\in\R_{>0}$ and $R_r\in\R_{>0}$ are the stator and rotor circuit resistances. The superscripts $d$ and $q$ refer to the vector component in the corresponding axis of the $dq$-reference frame, rotating at the synchronous speed $\omega_s$. The first two expressions in \eqref{eq:Vdsr} describe the stator voltage, whereas the latter two reflect the voltage circuit balance of a short-circuited rotor, hence $v_r=\mathbbl{0}_2$. Note that the slip frequency $\omega_\nu$ is involved in the last terms of \eqref{eq:Vdr}-\eqref{eq:Vqr}. Moreover, the stator and rotor flux linkages can be described as
\begin{subequations}
\label{eq:FsFr}
\begin{align}
    \psi_s &= L_s i_s + L_m i_r, \label{eq:Fs} \\ 
    \psi_r &= L_r i_r + L_m i_s, \label{eq:Fr} 
\end{align}
\end{subequations}
with $i_s = \left(i_s^d,i_s^q\right)\in\R^2$ and $i_r = \left( i_r^d,i_r^q \right)\in\R^2$ denoting the vectors comprising stator and rotor current components in different axes, and $L_s\in\R_{>0}$, $L_r\in\R_{>0}$, $L_m\in\R_{>0}$ being the stator, rotor and mutual inductance respectively. 

The electric power $p_e\in\R$ transferred between stator and rotor can now be expressed in terms of currents and flux linkages, either on the stator or on the rotor side as:
\begin{equation}
    p_e = \omega_s \frac{3}{2} \left( \psi_s^d i_s^q - \psi_s^q i_s^d \right) = \omega_s \frac{3}{2} \left( \psi_r^q i_r^d - \psi_r^d i_r^q \right), \label{eq:Pe}
\end{equation}
which yields the virtual electrical torque
\begin{equation}
    \tau_e = \frac{p_e}{\omega_s} = \frac{3}{2} \left( \psi_s^d i_s^q - \psi_s^q i_s^d \right) = \frac{3}{2} L_m \left( i_r^d i_s^q - i_r^q i_s^d \right). \label{eq:Te}
\end{equation}
It is worth emphasizing that the expression for $\tau_e\in\R$ in \eqref{eq:Te} is the same as for a synchronous machine~\cite{Kundur}. 

Considering the fact that the converter model will not involve a PLL (and hence the synchronous speed and slip are unknown to the controller), the presence of $\omega_s$ and $\omega_\nu$ terms in \eqref{eq:Vds}-\eqref{eq:Vqs} and \eqref{eq:Vdr}-\eqref{eq:Vqr}, respectively, poses an obstacle for the synchronization unit design. In other words, $\omega_s$ and $\omega_\nu$ are unknown variables and need to be computed internally based on available measurements. For that purpose, a field-oriented IM control first presented in \cite{DeDoncker1993} is employed. Considering that the direction of the $dq$-frame can arbitrarily be selected, we assume that in steady state the rotor flux is aligned with the $d$-axis, resulting in a simplified model with $\psi_r^q=0$. The above assumption resembles the ones used in conventional PLLs, where the calculation of the voltage angle is based on aligning the voltage vector with the $d$-axis of a synchronously-rotating reference-frame~\cite{Golestan2014}. 

According to the proposed approximation, \eqref{eq:Fr} is reformulated as
\begin{subequations}
\label{eq:Irdq}
\begin{align} 
    i_r^d &= \frac{\psi_r^d - L_m i_s^d}{L_r}, \label{eq:Ird}\\
    i_r^q &= - \frac{L_m}{L_r} i_s^q  \label{eq:Irq},
\end{align}
\end{subequations}
and the expressions for rotor voltage components in \eqref{eq:Vdr} and \eqref{eq:Vqr} can now be rewritten as
\begin{subequations}
\label{eq:Vdrqr2}
\begin{align} 
    0 &= R_r i_r^d + \dot{\psi}_r^d,  \label{eq:Vdr2} \\ 
    0 &= R_r i_r^q + \omega_\nu \psi_r^d.  \label{eq:Vqr2}
\end{align}
\end{subequations}
Substituting \eqref{eq:Ird} into \eqref{eq:Vdr2} yields
\begin{equation}
    \dot{\psi}_r^d = - R_r i_r^d = - \frac{R_r}{L_r} \left( \psi_r^d - L_m i_s^d \right), \label{eq:Vdr3_def}  
\end{equation}
which in frequency domain can be expressed as
\begin{equation}
    \psi_r^d(s) = \frac{R_r L_m}{R_r + s L_r} i_s^d = K_\psi(s) i_s^d. \label{eq:Vdr3}  
\end{equation}
Similarly, the slip frequency of the induction machine is computed by combining \eqref{eq:Irq} and \eqref{eq:Vqr2}:
\begin{equation}
    \omega_\nu = - \frac{R_r}{\psi_r^d} i_r^q =  \frac{R_r L_m}{L_r} \frac{i_s^q}{\psi_r^d}, \label{eq:Wslip_def}
\end{equation}
and subsequently substituting $\psi_r^d$ from \eqref{eq:Vdr3}, which gives
\begin{equation}
    \omega_\nu(s) = \left( \frac{R_r}{L_r} + s \right) \frac{i_s^q}{i_s^d} = K_\nu(s) \frac{i_s^q}{i_s^d}. \label{eq:Wslip}
\end{equation}
The expression \eqref{eq:Wslip} describes the dynamics of the slip frequency as a PD controller $K_\nu(s)$ applied to the ratio of $dq$-components of the stator current. As such, this term is clearly sensitive to the variations in grid frequency and machine power output. Nevertheless, in order to complete the PLL-less design, one needs to determine the synchronous speed. Having in mind that $\omega_s=\omega_r+\omega_\nu$, an exact estimation of the rotor's angular velocity is sufficient for achieving the targeted objective.
Let us observe the power balance of an induction machine via the swing equation and the mechanical dynamics of the rotor, given by:
\begin{equation} 
    J \dot{\omega}_r = \tau_m - \tau_e - \tau_d. \label{eq:swing}
\end{equation}
Here, $J\in\R_{>0}$ is the rotor's momentum of inertia, and $\tau_m\in\R$, $\tau_e\in\R$ and $\tau_d\in\R$ correspond to the mechanical, electrical and damping torque. By declaring $\Delta\omega_r\in\R$ as a deviation of ${\omega}_r$ from an initial (nominal) rotor speed\footnote{We denote the initial (nominal) rotor speed by $()^\star$ as it will later serve as an input setpoint to the VIM synchronization unit.} $\omega_0^\star\in\R_{\geq0}$, the expression \eqref{eq:swing} becomes
\begin{equation}     
    \Delta\dot{\omega}_r = \frac{1}{J} \left( \tau_m - \tau_e - \tau_d \right). \label{eq:swing1}
\end{equation}
By expressing all three torque components in \eqref{eq:swing1} as functions of converter input signals, one could finalize the closed-form VIM formulation. We elaborate on mathematical details and derivations in the remainder of this section.

The electrical torque component is defined in \eqref{eq:Te}, but can be further simplified by substituting the following expressions for stator flux linkage components:
\begin{subequations}
\label{eq:Fsdq}
\begin{align} 
    \psi_s^d &= \left( L_s - \frac{L_m^2}{L_r} \right) i_s^d + \frac{L_m}{L_r} \psi_r^d, \label{eq:Fsd} \\
    \psi_s^q &= \left( L_s - \frac{L_m^2}{L_r} \right) i_s^q, \label{eq:Fsq}
\end{align}
\end{subequations}
previously obtained from \eqref{eq:FsFr}. The electrical torque in time-domain is therefore of the form
\begin{equation}
    \tau_e = \frac{3}{2} \frac{L_m}{L_r} \psi_r^d i_s^q,  \label{eq:Te2_def}    
\end{equation}
whereas in frequency domain it yields
\begin{equation}
    \tau_e = \frac{3}{2} \frac{L_m}{L_r} K_\psi(s) i_s^d i_s^q = K_e(s) i_s^d i_s^q.  \label{eq:Te2} 
\end{equation}
In \eqref{eq:Te2}, $K_e(s)$ represents a first-order transfer function
\begin{equation}
    K_e(s) = \frac{3}{2} \frac{L_m}{L_r} K_\psi(s) = \frac{3}{2} \frac{R_r L_m^2}{R_r L_r + s L_r^2}, \label{eq:Ke}
\end{equation}
defined by the circuit parameters of the underlying induction machine, namely the rotor's resistance and reactance as well as the mutual inductance. 

On the other hand, the mechanical torque is determined by the machine's mechanical power output and the angular speed of the rotor. Assuming a lossless converter, the mechanical input power of an IM can be approximated by the output power measured at the converter terminal (denoted by $p_c\in\R$), and given by
\begin{equation}
\tau_m = \frac{p_m}{\omega_r} \approx \frac{p_c}{\omega_r}. \label{eq:tauM_def}
\end{equation}
Since the converter's terminal current and voltage measurements, corresponding to stator current and voltage of a virtual induction machine\footnote{In other words, $v_f\coloneqq v_s$ and $i_g\coloneqq i_s$, using the notation from Section~\ref{sec:modeling}.}, are available and actively employed in device-level control (i.e., inner loop control presented in the following section), the output active power can be expressed as $p_c\coloneqq v_f^\mathsf{T}i_g\coloneqq v_s^\mathsf{T}i_s$, therefore transforming \eqref{eq:tauM_def} into
\begin{equation}
\tau_m = \frac{v_s^di_s^d+v_s^qi_s^q}{\omega_0^\star + \Delta\omega_r}. \label{eq:tauM}
\end{equation}

Finally, the damping torque $\tau_d = D \Delta\omega_r$ is proportional to the rotor frequency deviation, which yields the following low-pass filter characteristic of the induction machine in frequency domain:
\begin{equation} 
    \Delta\omega_r = \frac{1}{Js + D} \left( \tau_m - \tau_e \right), \label{eq:swing2_old}
\end{equation}
where $D\in\R_{\geq0}$ denotes the damping constant. Substituting \eqref{eq:Te2} and \eqref{eq:tauM} into \eqref{eq:swing2_old} results in a closed-form expression for $\Delta\omega_r$ of the form:
\begin{equation} 
\Delta\omega_r = \frac{1}{Js + D} \left( \frac{v_s^di_s^d+v_s^qi_s^q}{\omega_0^\star + \Delta\omega_r} -  \frac{3}{2} \frac{R_r L_m^2}{R_r L_r + s L_r^2} i_s^d i_s^q \right),\hspace{-0.1cm} \label{eq:swing2}
\end{equation}
which corresponds to $\Delta\omega_r = F_r\left(u,p\right)$, with the measurement input vector $u\in\R^4$ and parameter vector $p\in\R_{>0}^6$ defined as
\begin{subequations}
\label{eq:u_p}
\begin{align} 
    u &= \big(v_s^d,v_s^q,i_s^d,i_s^q\big),\\
    p &= \left(J,D,R_r,L_r,L_m,\omega_0^\star\right).
\end{align}
\end{subequations}

Having obtained the desired analytical expressions for all frequency components, the synchronous speed can now be computed from the frequency slip $\omega_\nu$ in \eqref{eq:Wslip} and $\Delta\omega_r$ in \eqref{eq:swing2}, as follows:
\begin{equation}
    \begin{split}
        \omega_s &= \omega_r + \omega_\nu \\
        &= \underbrace{\omega_0^\star + F_r(u,p)}_{\omega_r} + \underbrace{\left( \frac{R_r}{L_r} + s \right) \frac{i_s^q}{i_s^d}}_{\omega_\nu}\\ 
        &\eqqcolon F_s(u,p). \label{eq:ws}
    \end{split}
\end{equation}
Similarly to any PLL, the angle reference is determined by integrating the frequency signal, i.e.,  $\dot{\theta}_s=\omega_s$. The resulting expression in \eqref{eq:ws} clearly shows that the closed-loop estimator $F_s(u,p)$ emulates the synchronous speed and thus the synchronization properties of an IM based solely on the voltage and current measurements $v_f$ and $i_g$ at the converter terminal, therefore entirely replacing the conventional PLL-based synchronization. 

On the downside, the difference between the true synchronous speed and initial rotor speed setpoint can have an impact on synchronization accuracy. In particular, a proper selection of $\omega_0^\star$ prior to the grid connection of the VSC reduces $\Delta\omega_r$ and the subsequent transients. Nevertheless, it is reasonable to assume that the VSC is connected to the grid during steady-state operation. Furthermore, even if the system is not in steady-state, any reasonable $\omega_0^\star$ guess will still allow the VIM to synchronize at the cost of some minor transients. A sensitivity analysis addressing the underlying phenomena is provided in Section~\ref{subsec:startup}. 
Another potential drawback of the frequency estimator \eqref{eq:ws} is the fact that the slip frequency $\omega_\nu$ is computed using a PD controller imposed on the quotient of the current components in $dq$-frame. On the one hand, the derivative control is sensitive to fast signal changes. As the quotient $i_s^q/i_s^d$ can experience high oscillations during transients, the PD controller might be prone to overshoots and even instability. On the other hand, the given input-output structure of the PD controller might lead to an index-2 DAE form\footnote{Index is a notion used in the theory of DAEs for measuring the distance from a DAE to its related ODE (i.e., the number of  differentiations needed to obtain the ODE form). It is a non-negative integer that provides useful information about the mathematical structure and potential complications in the analysis and the numerical solution of the DAE. In general, the higher the index of a DAE, the more difficulties one can expect for obtaining its numerical solution~\cite{Petzold1995}.}, which in turn increases the computational burden and imposes restrictions on the selection of the DAE solver. The aspects of DAE formulation will be discussed in detail in Section~\ref{subsec:ctrl_dae}.

\begin{figure*}[!t]
    \centering
    \scalebox{0.725}{\includegraphics[]{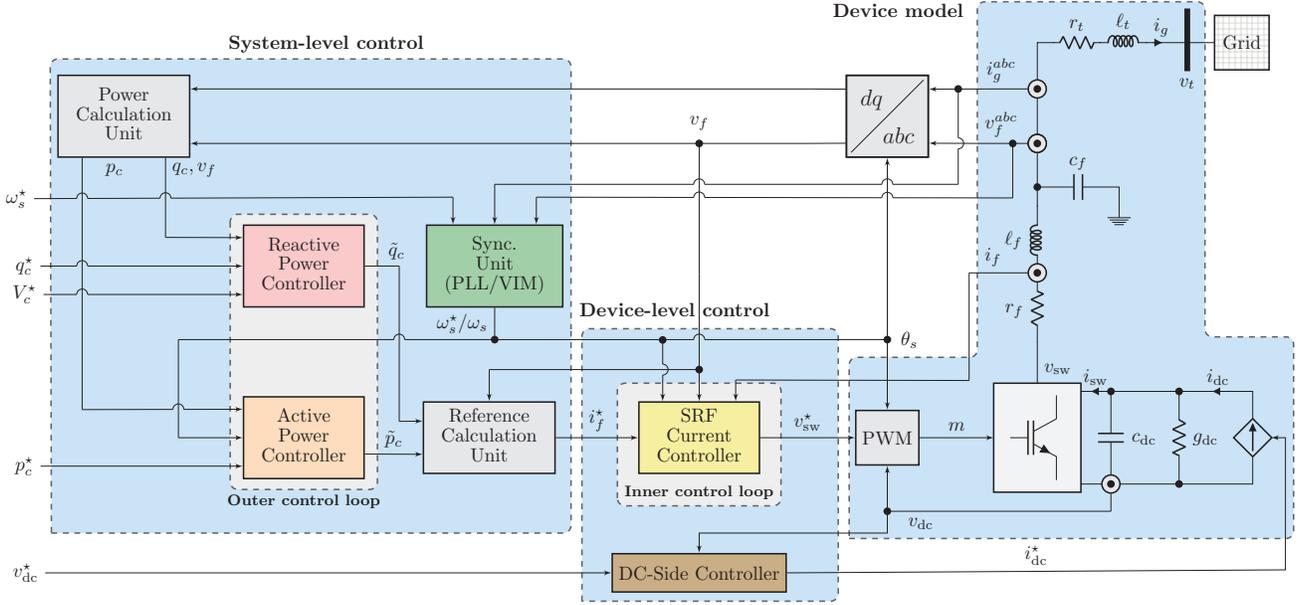}}
    \caption{General configuration of the implemented grid-following VSC control structure.}
    \label{fig:vsc_ctrl}
\end{figure*}
\section{VSC Control \& Synchronization} \label{sec:modeling}
An overview of the prevalent control architecture for two-level grid-following power converters is shown in Fig.~\ref{fig:vsc_ctrl}. The power converter model considered in this study is composed of a DC-link capacitor, an averaged switching block which modulates the DC-capacitor voltage $v_\mathrm{dc} \in \R_{>0}$ into an AC voltage $v_{\mathrm{sw}}\in\R^2$, and an output filter. Furthermore, we assume that the DC-source current $i_\mathrm{dc}\in\R_{>0}$ is supplied by a controllable source, in the form of energy storage or curtailed renewable generation, and can be used as a control input. In the considered control structure, an outer \textit{system-level} control provides a reference for the converter's terminal current that is subsequently tracked by a \textit{device-level} controller. In the following, the individual control blocks depicted in Fig.~\ref{fig:vsc_ctrl} are discussed together with the respective DAE formulations.

\subsection{System-Level Control for the Grid-Following Mode} \label{subsec:sys_level}
In the system-level control of grid-following VSCs, the input measurements $y_s=(v_f,p_c,q_c)\in\R^4$ are commonly the output voltage $v_f \in \R^2$ and the instantaneous active and reactive power given by
\begin{equation}
    p_c\coloneqq v^\mathsf{T}_f i_g, \quad q_c \coloneqq v^\mathsf{T}_f j^\mathsf{T} i_g, \label{eq:measurements2}
\end{equation} 
where $j \in\R^{2\times2}$ is the $\ang{90}$ rotation matrix, and $i_g \in \R^2$ denotes the converter current injection into the system. Moreover, a synchronization device - a PLL or a VIM - estimates the phase angle $\theta_s \in [-\pi,\pi)$ of the voltage $v_f$ as well as the synchronous (grid) frequency $\omega_s\in\R_{>0}$ at the point of common coupling, and provides them as reference $(\theta_s,\omega_s)$ to the device-level control. 
%In addition, the outer control loop is used to calculate the current reference $i_f^\star\in\R^2$ based on the mismatch between measured signals $y_s$ and prescribed setpoints $u_s=(p_c^\star,q_c^\star,\omega_s^\star,V_c^\star)\in\R^4$. 
Having determined the synchronous angle and frequency, the outer control loop subsequently computes the current reference $i_f^\star\in\R^2$, by employing frequency and voltage droop control $(R_c^p$, $R_c^q)\in\R_{\geq0}^2$ in combination with integral controllers $K_{I,f}^d\in\R_{>0}$ and $K_{I,f}^q\in\R_{>0}$. More precisely, the outer control loop, described by internal state variables $(\tilde{p}_c,\tilde{q}_c)\in\R^2$, regulates the power output $(p_c,q_c)$ to its respective setpoint $(p_c^\star,q_c^\star)\in\R^2$, as follows:
\begin{subequations}\label{eq:sysLevel_follow}
\begin{align}
    \dot{\tilde{p}}_c &= K_{I,f}^d \left(p_c^\star - p_c - R_c^p\left(\omega_s-\omega_s^\star\right) \right),\\
    \dot{\tilde{q}}_c &= K_{I,f}^q \left(q_c^\star - q_c - R_c^q\left(\norm{v_f} - V_c^\star \right)\right).
\end{align}
\end{subequations}
Due to the $P-f$ and $Q-V$ droop characteristics, the active power reference is adjusted in response to a deviation of the measured frequency $\omega_s$ with respect to the frequency setpoint $\omega_s^\star\in\R_{>0}$, whereas the reactive power reference is modified according to the mismatch between the magnitude of the output voltage $\norm{v_f}\in\R_{\geq0}$ and the converter voltage setpoint $V_c^\star\in\R_{>0}$. 
%Hence, the internal state vector of the system-level controller is $x_s=(\theta_s,\varepsilon,\tilde{p}_c,\tilde{q}_c)$.

The computed active and reactive power references are then transformed into the corresponding current reference signal $i_f^\star$ by adjusting the power references $(\tilde{p},\tilde{q})$ based on output voltage measurement such that the converter's power output is kept constant:
\begin{equation} \label{1.2eq:ref_constP}
    {i_f^\star}^d=\frac{v_f^d \tilde{p}_c + v_f^q \tilde{q}_c}{\norm{v_f}}, \quad   {i_f^\star}^q=\frac{v_f^q \tilde{p}_c - v_f^d \tilde{q}_c}{\norm{v_f}}.
\end{equation}

\subsection{Grid Synchronization: PLL vs. VIM} \label{subsec:sync_pll_vim}

\begin{figure}[!t]
\centering
\begin{minipage}{0.485\textwidth}
    \centering
    \scalebox{0.75}{\includegraphics[]{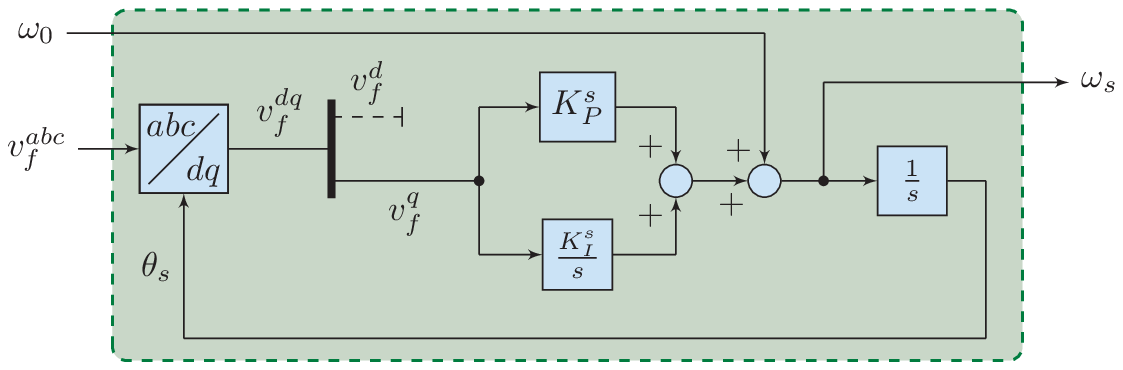}} \\
    {\footnotesize (a)}   
    \vspace{0.5em}
\end{minipage}
\begin{minipage}{0.485\textwidth}
    \centering
    \scalebox{0.75}{\includegraphics[]{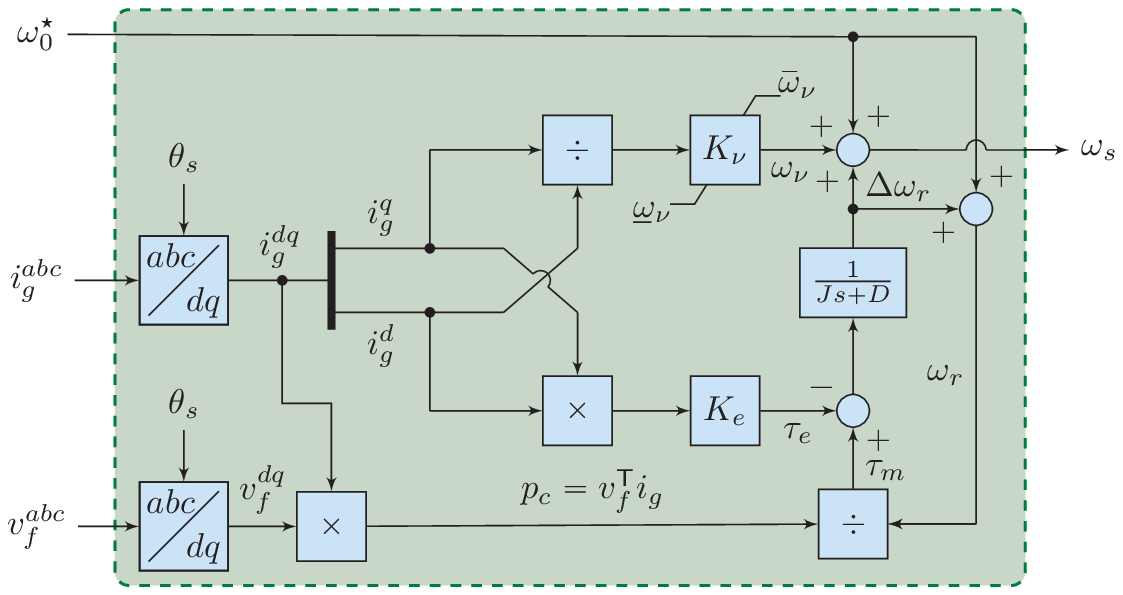}} \\
    {\footnotesize (b)}   
\end{minipage}
\caption{\label{fig:sync_blocks}Different synchronization units of a power converter: (a) phase-locked loop \eqref{eq:pll_basic}; (b) virtual induction machine \eqref{eq:ws}.}
\vspace{-0.35cm}
\end{figure}

The two considered synchronization methods are analyzed in this subsection. For comparison and clarity, the control-block implementation of both synchronization units (i.e., a PLL and a VIM) is depicted in Fig.~\ref{fig:sync_blocks}. Notice that in the proposed converter control scheme presented in Fig.~\ref{fig:vsc_ctrl}, the VIM solely replaces the PLL as a new synchronization unit (the green block). Therefore, active and reactive power tracking functionalities of the grid-following control remain unchanged and are governed by the active and reactive power controllers within the outer control loop.

The most common PLL implementation is a so-called type-2 Synchronously-rotating Reference Frame\footnote{The synchronous reference frame is a concept related to the $dq$-transformation, where rotating vectors are described with respect to a rotating coordinate system \cite{SyncReview2020}. When the coordinate system is rotating at synchronous speed, it is typically referred to as the synchronously rotating reference frame.} (SRF) PLL~\cite{Chung2000}, which acts as an observer and tracks the synchronous speed by measuring the stationary output voltage $v_f$, transforming it into an internal $dq$-SRF, and passing it through a PI-controller $(K_P^s,K_I^s)\in\R^2_{>0}$ that acts on the phase angle difference:
\begin{equation}
    \omega_s = \omega_0 + \left(K_P^s + \frac{K_I^s}{s} \right) v_f^q, \label{eq:pll_basic}
\end{equation}
with $\omega_0\in\R_{>0}$ denoting the nominal angular velocity. The synchronization is achieved by aligning the $d$-axis of the internal SRF with the stationary $abc$-frame and diminishing the $q$-component, as described in~\cite{Chung2000,UrosGM} and illustrated in Fig.~\ref{fig:sync_blocks}. In other words, the combined Clarke and Park transformation within the PLL introduces an SRF into the system, which aims to align with the grid SRF. Representation of the PLL unit in DAE form is straightforward to obtain and can be written as\cite{Misyris2019}:
\begin{subequations} \label{eq:PLL}
\begin{align}
    \dot{\varepsilon} &= v_f^q, \label{1.2eq:epsilon_pll}\\
    \omega_s &= \omega_0 + K_P^s v_f^q + K_I^s \varepsilon, \label{1.2eq:w_pll}\\
    \dot{\theta}_s &= \omega_b\omega_s, \label{1.2eq:theta_pll}
\end{align}
\end{subequations}
with $\varepsilon\in\R$ denoting the integrator state.

The VIM control design illustrated in Fig.~\ref{fig:sync_blocks} is based on \eqref{eq:ws}. Similarly to other synchronization methods including the PLL, the unknown grid frequency can be obtained by simply measuring the three-phase current and voltage at the filter output $(i_g^{abc},v_f^{abc})$. Note that the VIM also operates in an SRF, defined by the internally computed synchronization angle $\theta_s$. As previously explained in Section~\ref{sec:vim_strategy}, another necessary input for the controller is the initial rotor frequency $\omega_0^\star$, which determines the VIM's oscillation level at start-up. However, the requirement for the initial guess of $\omega_0^\star$ is not very strict, as it should only be ``close enough'' to the synchronous speed and subsequently let the emulated induction machine bring the VSC to synchronism. Moreover, in order to cope with potential frequency slip spikes during transients, induced by the PD control $K_\nu(s)$ acting on the quotient $i_g^q/i_g^d$ in \eqref{eq:Wslip}, the frequency slip is constrained by the saturation limits $\omega_\nu\in\left[\ushort{\omega}_\nu,\widebar{\omega}_\nu\right]$.

In addition, Figure~\ref{fig:sync_blocks} sheds light on two potential drawbacks of the VIM-based synchronization unit. Firstly, the structure of the VIM unit is more complex and relies on more tuning parameters compared to PLL. Secondly, while PLL relies only on the terminal voltage measurement, VIM in addition requires the three-phase current measurement at the VSC output, therefore potentially being more prone to measurement delays and errors.

\subsection{System-Level Control for the Grid-Forming Mode} \label{subsec:sys_level_form}
For comparison purposes, in this section we introduce the grid-forming control mode. In contrast to the previously presented grid-following operation mode, grid-forming control achieves synchronization by measuring power imbalance and is inspired by the traditional primary frequency control of synchronous machines. More precisely, by appropriately adjusting the individual droop factors, it enables self-synchronization to the power grid and power sharing in proportion to the converter rating, while using only locally available measurements. Droop controlled grid-forming control mode is given by \cite{SimpsonPorco2017,UrosStability2021}:  
\begin{subequations} \label{1.2eq:powersync}
\begin{align} 
	\omega_c &= \omega_s^\star + {R_c^p \lambda_z(s) \left(p_c^\star - p_c\right)},\\
	v_c^d &= V_c^\star + {R_c^q \lambda_z(s) \left(q_c^\star - q_c\right)},
\end{align}
\end{subequations}
with $R_c^p\in\R_{\geq0}$ and $R_c^q\in\R_{\geq 0}$ being the droop gains, $\lambda_z(s)=\frac{\omega_z}{\omega_z + s}$ representing the low-pass filter applied to the power measurements, $\omega_z\in\R_{>0}$ denoting its cut-off frequency, $\dot{\theta}_c = \omega_b\omega_c$ and $v_c^q=0$. It is worth noting that, unlike for the grid-following operation mode, the angle $\theta_c$, frequency $\omega_c\in\R$, and voltage $v_c\in\R^2$ are internal states of the control and no synchronization unit is needed. Only the active active and reactive power injections $p_c$ and $q_c$ are measured for purposes of the system-level control.  
% Therefore, the control state vector can be defined as $x_s^\mathrm{form}=(\theta_c,\tilde{\omega}_c,\tilde{v}_c)$, where
% \begin{subequations}
% \begin{align}
%     \dot{\tilde{\omega}}_c &= -\omega_z\tilde{\omega}_c + R_c^p\omega_z(p_c^\star-p_c), \\
%     \dot{\tilde{v}}_c &= -\omega_z\tilde{v}_c + R_c^q\omega_z(q_c^\star-q_c).
% \end{align}
% \end{subequations}

\subsection{Device-Level Control} \label{subsec:device_level_ctrl}
This control layer provides both AC and DC-side reference signals for the VSC device. In case of the grid-forming control mode, the AC-side controller consists of a cascade of voltage and current controllers (also called inner control loops) computing a voltage reference for the modulation unit. In contrast, the system-level controller of a grid-following converter consists only of a current PI controller tracking the filter current reference, as described in Fig.~\ref{fig:vsc_ctrl}. Finally, the DC voltage $v_\mathrm{dc}$ in both cases is controlled through the DC current source by a PI controller. Fore more details on the device-level control we refer the reader to \cite{UrosStability2021}.
% Unlike the widely used virtual synchronous machine control \cite{DArco2013_2,Zhong2011}, and in contrast to the claims raised in \cite{Ashabani2016}, the inertia and damping in \eqref{eq:swing2} will not be reflected in the converter's frequency response and oscillation damping performance. They do, however, contribute to a more robust and resilient frequency estimation technique, as will be later shown in Section~\ref{sec:5}. 

\subsection{Virtual Induction Machine DAE Formulation} \label{subsec:ctrl_dae}
The DAE formulation of a PLL synchronization unit has previously been presented in Section~\ref{subsec:sync_pll_vim}. It was shown that the PLL controller \eqref{eq:pll_basic} can be expressed as a second-order dynamic system with state vector $x_\mathrm{pll}\coloneqq\left(\varepsilon,\theta_s\right)\in\R^2$ and algebraic output vector $y_\mathrm{pll}\coloneqq\omega_s$.

In contrast, obtaining an appropriate (i.e., index-1) DAE representation of the VIM controller is not straightforward due to the aforementioned issues pertaining to the computation of the slip frequency in \eqref{eq:Wslip}, as well as the fact that the electrical torque is described by a first-order transfer function $K_e(s)$ applied to the product of state variables $i_g^d$ and $i_g^q$ in \eqref{eq:Te2}. As such, the existing DAE model is incomplete since the number of algebraic equations does not correspond to the number of algebraic variables. More precisely, the algebraic equation describing the slip frequency is missing and must be derived by transforming \eqref{eq:Wslip} accordingly. 

We tackle this issue by introducing a new variable $\varphi\in\R$ such that
\begin{equation}
    \varphi \coloneqq \frac{\diff}{\diff\!t}\!\left(\frac{i_g^q}{i_g^d}\right). \label{eq:quot}
\end{equation}
Moreover, we define $\varphi_d\coloneqq\dot{i}_g^d$ and $\varphi_q\coloneqq\dot{i}_g^q$, and apply the quotient rule to \eqref{eq:quot}, which yields
\begin{equation}
     \varphi = \frac{\dot{i}_g^q i_g^d - i_g^q \dot{i}_g^d}{\left(i_g^d\right)^2} = \frac{1}{i_g^d} \varphi_q - \frac{i_g^q}{\left(i_g^d\right)^2} \varphi_d. \label{eq:quot2}
\end{equation}
The terms $(\varphi_d,\varphi_q)\in\R^2$ can be related to the differential equation describing the dynamics of the current flowing through the transformer at the converter output:
\begin{equation}
    \dot{i}_g=\frac{\omega_b}{\ell_t}(v_f-v_t)-\left(\frac{r_t}{\ell_t}\omega_b+j\omega_b \omega_s\right)i_g, \label{eq:elSys3}
\end{equation}
where $r_t\in\R_{>0}$ and $\ell_t\in\R_{>0}$ denote the transformer's resistance and inductance, and $v_t\in\R^2$ is the voltage at the connection terminal.
Redefining \eqref{eq:elSys3} in SI yields
\begin{subequations}
\label{eq:phidq}
\begin{align}
    \varphi_d&=\frac{\omega_b I_b}{\ell_t}\left(v_f^d-v_t^d\right)- I_b \left(\frac{r_t}{\ell_t}\omega_b i_g^d + \omega_b\omega_s i_g^q \right), \label{3.1eq:phid} \\
    \varphi_q&=\frac{\omega_b I_b}{\ell_t}\left(v_f^q-v_t^q\right)- I_b \left(\frac{r_t}{\ell_t}\omega_b i_g^q + \omega_b\omega_s i_g^d \right). \label{3.1eq:phiq} 
    \end{align}
\end{subequations}
Here, $\omega_b\in\R_{>0}$ and $I_b\in\R_{>0}$ denote the base angular velocity and current used for conversion between the per unit and SI. The rest of the notation is adopted from \eqref{eq:elSys3}. By rewriting \eqref{eq:Te2} and \eqref{eq:swing2} as
\begin{subequations}
\label{eq:diff_new}
\begin{align}
    \dot{\tau}_e &= -\frac{R_r}{L_r} \tau_e + \frac{3 R_r L_m^2}{2 L_r^2} i_g^d i_g^q,  \label{3.1eq:Te_new} \\
    \Delta\dot{\omega}_r &= \frac{1}{J} \left( \frac{v_f^\mathsf{T}i_g}{\omega_0^\star + \Delta\omega_r} - \tau_e \right) - \frac{D}{J} \Delta\omega_r, \label{eq:swing_new}
\end{align}
\end{subequations}
and transforming \eqref{eq:Wslip} and \eqref{eq:ws} respectively into
\begin{subequations}
\label{eq:ws_slip_new}
\begin{align}
    \tilde{\omega}_\nu = \frac{R_r}{L_r} \frac{i_g^q}{i_g^d} + \varphi, \label{3.1eq:Wslip_new}\\
    \omega_s = \omega_0^\star + \Delta\omega_r + \tilde{\omega}_\nu, \label{3.1eq:ws_new}
\end{align}
\end{subequations}
we obtain an index-1 DAE form comprising \eqref{eq:diff_new} as differential equations and \eqref{eq:quot2}-\eqref{eq:phidq}, and \eqref{eq:ws_slip_new} as algebraic equations\footnote{The state vectors $v_f$, $i_g$ and $v_t$ are included in the dynamical model of the converter control and hence do not contribute to the model order of the synchronization unit.}. The state vector representing the VIM dynamics is thus $x_\mathrm{vim}=\left(\tau_e,\Delta\omega_r\right)\in\R^2$, and is of the same order as the PLL controller, whereas the vector of algebraic variables consists of $y_\mathrm{vim}=\left(\varphi,\varphi_d,\varphi_q,\tilde{\omega}_\nu,\omega_s\right)\in\R^5$. The newly defined variable $\tilde{\omega}_\nu\in\R$ represents the unsaturated frequency slip. 

In order to include the frequency slip saturation limits into the DAE model, we employ the well-known expressions for the minimum and maximum of two variables:
\begin{subequations}
\label{eq:minmax}
\begin{align}
    \min\{a,b\} &= \frac{a+b-\left|b-a\right|}{2},\\
    \max\{a,b\} &= \frac{a+b+\left|b-a\right|}{2},
\end{align}
\end{subequations}
and re-declare $\omega_\nu\in\left[\ushort{\omega}_\nu,\widebar{\omega}_\nu\right]$ as the saturated slip counterpart of $\tilde{\omega}_\nu$ determined by the following algebraic equation:
\begin{equation}
\begin{split}
    \omega_\nu = \frac{1}{2} \Bigl(& \frac{1}{2} \bigl( \tilde{\omega}_\nu + \ushort{\omega}_\nu + | \ushort{\omega}_\nu - \tilde{\omega}_\nu | \bigr) + \widebar{\omega}_\nu - \\ \bigl|& \widebar{\omega}_\nu - \frac{1}{2} ( \tilde{\omega}_\nu + \ushort{\omega}_\nu + | \ushort{\omega}_\nu - \tilde{\omega}_\nu |)\bigr| \Bigr). \label{eq:slip_sat}
\end{split}
\end{equation}
The derivation of \eqref{eq:slip_sat} is given in Appendix~\ref{sec:app_3.1}. 

By including $\omega_\nu$ into $y_\mathrm{vim}$ we complete the DAE formulation of the VIM controller. Combining it with the internal dynamics of converter's system- and device-level control, as well as with the network-side dynamics pertaining to the device model representation, results in a $16^\mathrm{th}$-order grid-following converter model for both the PLL and VIM-based synchronization principle.

\section{Simulation Results} \label{3.1sec:sim_results}
\begin{figure}[!b]
    \centering
    \includegraphics[scale=0.775]{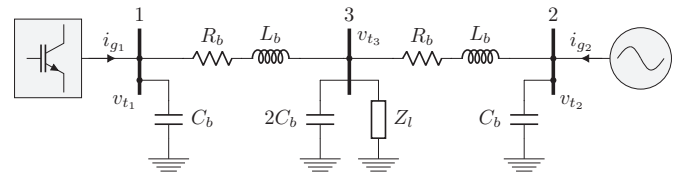}
    \caption{Schematic of the investigated 3-bus system, where a VSC is connected at bus 1, a stiff grid equivalent at bus 2, and an impedance load at bus 3. The line parameters are: $R_b=0.014\,\mathrm{p.u.},L_b=0.14\,\mathrm{p.u.},C_b=0.074\,\mathrm{p.u.}$}
    \label{fig:3bus_system}
\end{figure}
\subsection{System Setup and VIM Parameterization} \label{3.1subsec:setup}
In this section, the performance of the proposed synchronization scheme is studied for various transient scenarios using a detailed EMT model of a simple 3-bus system presented in Fig.~\ref{fig:3bus_system}, composed of a grid-following VSC, a stiff grid, and an $RL$ load, which are interconnected with transmission lines modeled as $\pi$-sections. The parameters used for the VIM synchronization unit, represented by model \eqref{eq:swing2}, have been obtained from a physical induction generator of a \SI{1.5}{\mega\watt} type-1 wind turbine, with the most relevant parameters for the VIM design listed in Table~\ref{tab:param}. The model is implemented in \textsc{Matlab} Simulink.

%Understandably, the VIM synchronization performance is highly dependent on the selection of model parameters, in particular the parameters of the equivalent physical induction machine. The response is particularly sensitive to the rotor resistance and inductance, but also to the mutual inductance included in the transfer function $K_e(s)$.} Additionally, proper values for the moment of inertia and damping are crucial for the dynamics of the rotor frequency, which in turn affects the sinusoidal nature of the voltage and current at the converter terminal.
%Note that the VIM input frequency is set to $f_0^\star=\SI{50}{\hertz}$. 
\begin{table}[!t]
\renewcommand{\arraystretch}{1.5}
\caption{Simulation parameters of the proposed VIM controller.}
\label{tab:param}
\noindent
\centering
    \begin{minipage}{\linewidth} %Use the minipage environment to footnote tables
    \renewcommand\footnoterule{\vspace*{-5pt}} %to remove the horizontal rule above the table footnote
    \begin{center}
    \scalebox{0.95}{%
        \begin{tabular}{ c || c | c }
            \toprule
            \textbf{Parameter} & \textbf{Symbol} & \textbf{Value} \\ 
            \cline{1-3}
            Nominal rated power & $P_n$ & $\SI{1.5}{\mega\watt}$\\
            Total moment of inertia\footnote{Corresponding to the normalized inertia constant $H=5\,\mathrm{s}$.} & $J$ & $152.14 \, \mathrm{kg\,m^2}$\\
            Damping constant & $D$ & $10 \, \mathrm{N\,m/s}$\\
            \arrayrulecolor{black!30}\hline
            Rotor resistance & $R_r$ & $0.0005 \, \mathrm{p.u.}$\\            
            Rotor inductance & $L_r$ & $0.05 \, \mathrm{p.u.}$\\
            Mutual inductance & $L_m$ & $0.6 \, \mathrm{p.u.}$\\        
            \arrayrulecolor{black!30}\hline
            Initial frequency setpoint\footnote{Corresponding to $\omega_0^\star=314.16\,\mathrm{rad/s}$.} & $f_0^\star$ & $50 \, \mathrm{Hz}$\\
            \arrayrulecolor{black}\bottomrule
        \end{tabular}
    }
        \end{center}
    \end{minipage}
\end{table}
The dynamics of the frequency slip are described by a PD controller $K_\nu(s)$ in \eqref{eq:Wslip}, with the proportional gain $K_{\nu}^P = R_r / L_r$ given by the IM parameters and a unity derivative gain $K_{\nu}^D = 1$. Nevertheless, such high value of the derivative gain can destabilize the converter control during transients. This problem is overcome by re-tuning the PD controller via the Ziegler-Nichols method \cite{Ziegler1993}, i.e., determining the optimal $K_{\nu}^D$ component while assuming the existing proportional gain $K_{\nu}^P$, which results in $K_{\nu}^D = 0.001$. The parameters of the system-level control (i.e., droop and integral gains included in the power control) and the device-level control (i.e., PI gains of the current controllers) have been adopted from \cite{UrosStability2021} in order to test the plug-n-play properties of the VIM.
\begin{figure}[!b] % <--- only p
    \centering
    \begin{minipage}{0.45\textwidth}
	\centering
	\scalebox{1.125}{\includegraphics[]{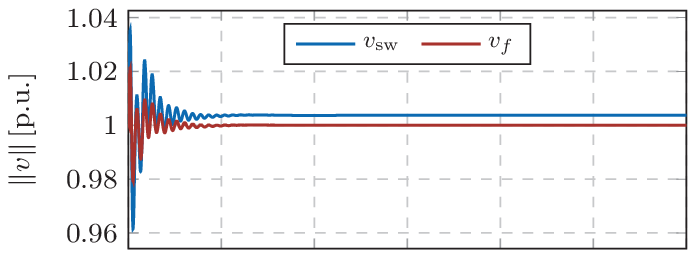}}\\
\end{minipage}
\begin{minipage}{0.45\textwidth}    
	\centering
	\scalebox{1.125}{\includegraphics[]{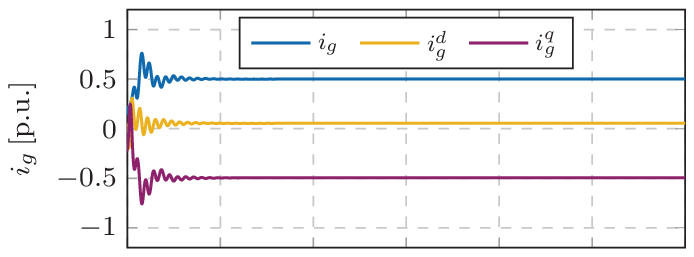}}\\
\end{minipage}
\begin{minipage}{0.45\textwidth}    
	\centering
	\scalebox{1.125}{\includegraphics[]{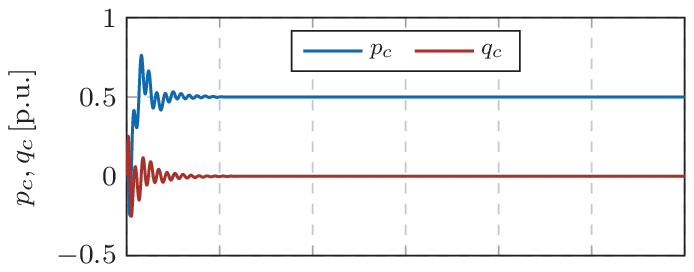}}\\
\end{minipage}
\begin{minipage}{0.45\textwidth}
	\centering
	\scalebox{1.125}{\includegraphics[]{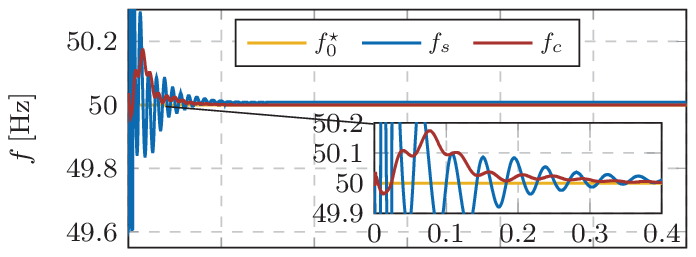}}\\
\end{minipage}
\begin{minipage}{0.45\textwidth}    
	\centering
	\scalebox{1.125}{\includegraphics[]{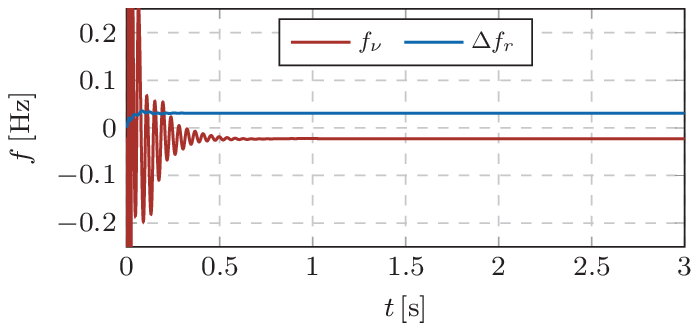}}\\
\end{minipage}
\caption{\label{fig:sync}Transient behavior of the VIM-based unit during start-up: (i) voltages; (ii) current and its $dq$-components; (iii) active and reactive power; (iv) frequencies; and (v) VIM frequency terms. }
\end{figure}
In the rest of this section, we first focus on real-time VSC operation events such as start-up and synchronization, response to setpoint variation (i.e., voltage and power reference tracking), as well as the impact of the initial rotor speed estimate $\omega_0^\star$ on the converter's synchronization process with the grid. Then, we analyze its behavior under a short-circuit fault. Subsequently, we compare grid-forming, PLL, and VIM-based grid-following VSCs under islanding (i.e., disconnection from the grid) and frequency disturbance events.

\subsection{Real-time Operation of VIM-synchronized VSCs} \label{subsec:startup}
First, the connection and synchronization of a VIM-based converter to the grid is studied. The VSC is connected to the grid at $t=0\,\mathrm{s}$, with the initial input frequency $f_0^\star = 50 \, \mathrm{Hz}$ set equal to the grid frequency. The voltage reference is initialized at $V_c^\star = 1 \, \mathrm{p.u.}$, whereas the active and reactive power setpoints are set to $p_c^\star = 0.5 \, \mathrm{p.u.}$ and  $q_c^\star = 0 \, \mathrm{p.u.}$, respectively.
The transient response illustrated in Fig.~\ref{fig:sync} confirms the soft-start and self-synchronization capabilities of the VIM as well as an adequate oscillation damping characteristic. The setpoints are correctly tracked and the voltage and current overshoots during start-up are acceptable. Furthermore, the start-up overshoots can be avoided if the VSC is slowly ramped-up from the zero power setpoint. Note that the initial overcurrents are in accordance with the characteristic response of an induction generator, but can also be assigned to numerical initialization of the model. The initial transients are better understood by observing the estimated synchronous frequency $f_s$ and its time-variant components $f_\nu$ and $\Delta f_r$. The frequency slip is very volatile during the first \SI{300}{\milli\second}, unlike the rotor's frequency dynamics term $\Delta f_r$, which can be associated with two aspects: (i) the frequency slip is proportional to the quotient $i_g^q/i_g^d$, which can reach very high values when $i_g^d \approx 0$; (ii) $K_\nu(s)$ behaves as a PD controller, with its derivative actions ($K_\nu^D$) being mostly utilized throughout the first \SI{300}{\milli\second} of the start-up. After \SI{500}{\milli\second}, both frequency components stabilize and the synchronous and converter output frequency reach a steady state value of $f_s\approx \SI{50.008}{\hertz}$. We found that the operation of the VIM-based unit with a negative active power setpoint is also feasible, with synchronization properties and responses resembling the ones presented in this case study. Thus, both load and generator operation modes are viable for VIM-synchronized converters. 
\begin{figure}[!t]
	\centering
	\begin{minipage}{0.45\textwidth}
		\centering
		\hspace{-0.265cm}
		\scalebox{1.125}{\includegraphics[]{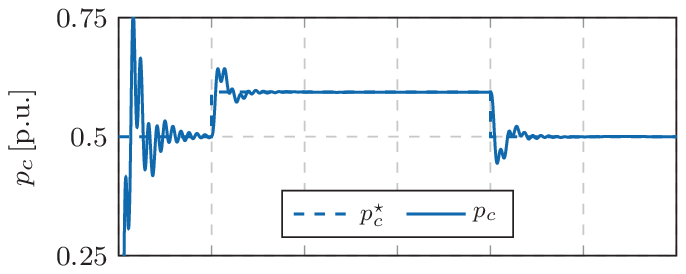}}\\
		\vspace{-0.1cm}  
	\end{minipage}
	\begin{minipage}{0.45\textwidth}    
		\centering
		\hspace{-0.15cm}
		\scalebox{1.125}{\includegraphics[]{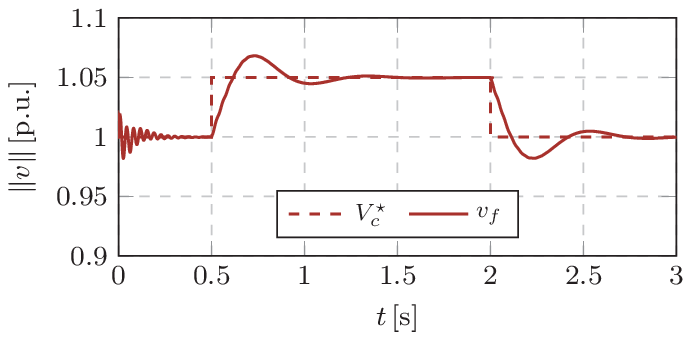}}\\
	\end{minipage}
	\caption{\label{fig:ref_var}VIM-based converter response to the variation of controller setpoints: (a)  Variation of the active power setpoint. (b) Variation of the voltage setpoint.}
\end{figure}

Another important aspect of control performance is the reference tracking, i.e., converter's ability to follow sudden changes in voltage and power setpoints. Both scenarios are simulated independently, with setpoint changes occurring at $t=\SI{0.5}{\second}$ in each case. The voltage reference exhibits a step increase of \SI{5}{\percent}, whereas the active power reference increases by \SI{20}{\percent}. Both step changes last for \SI{1.5}{\second} before setpoints returning to their original values. 
The results depicted in Fig.~\ref{fig:ref_var} indicate that power and voltage reference tracking is achieved within reasonable time, as both the output voltage (i.e., the voltage $v_f$ after the filter) and active power follow closely the respective setpoints. This is an expected outcome, as the reference tracking capability comes from the proper design of system- and device-level controls, which remain the same for both PLL and VIM-synchronized VSCs. Nevertheless, an inefficient synchronization device would have deteriorated the performance, which is clearly not the case for the VIM. A somewhat delayed response and excessive overshoot in the voltage output is solely an artifact of the employed PI tuning of the inner control loops, since the selected tuning favors the power tracking over the voltage tracking, and can easily be addressed with a different set of PI control gains.

\begin{figure}[!t]
	\centering
	\begin{minipage}{0.45\textwidth}
		\centering
		\scalebox{1.125}{\includegraphics[]{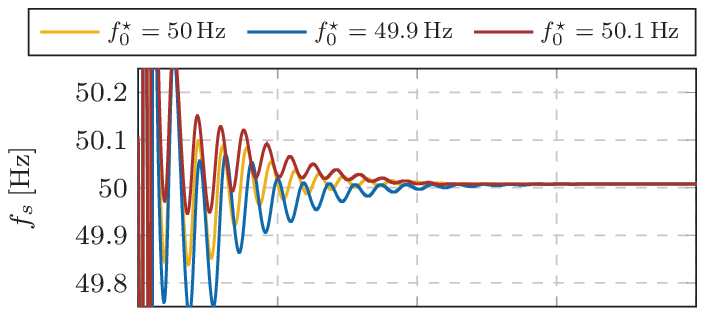}}\\
		\vspace{-0.05cm}  
	\end{minipage}
	\begin{minipage}{0.45\textwidth}    
		\centering
		\scalebox{1.125}{\includegraphics[]{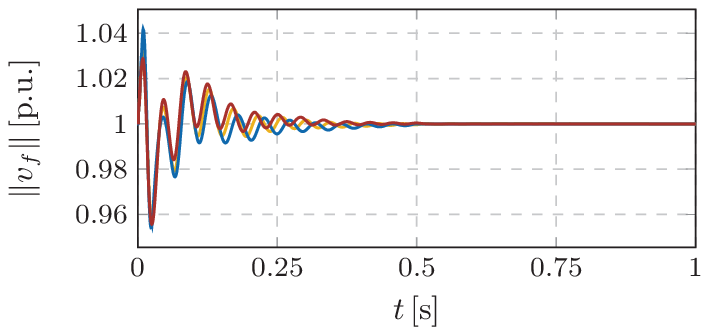}}\\
		\vspace{0.5em}      
	\end{minipage}
	\caption{\label{3.1fig:f0_var}Impact of the initial rotor frequency term on the synchronization process of a VIM-based VSC during start-up: synchronous VIM frequency (top) and  converter output voltage (bottom).}
\end{figure}

Let us recall that one of the control inputs to the VIM is the so-called ``estimated'' initial rotor frequency $f_0^\star$. Previous examples have shown that under the $f_s\approx f_0^\star$ condition the system experiences a satisfactory performance with good synchronization and damping properties. However, having knowledge of the exact grid frequency prior to the connection of the VSC might not be achievable in real-world applications. Thus, the impact of an inaccurate $f_0^\star$ guess on converter synchronization is investigated by studying the frequency and voltage response for $f_0^\star=\SI{49.9}{\hertz}$ and $f_0^\star=\SI{50.1}{\hertz}$, and comparing it against the results presented in Fig.~\ref{fig:sync} for the ``ideal'' case where $f_0^\star=\SI{50}{\hertz}$. The analysis is focused on the first second of the response after start-up, with the frequency and voltage response presented in Fig.~\ref{3.1fig:f0_var}. We can conclude that the synchronization is successfully achieved within \SI{500}{\milli\second} for all three scenarios, with no distinctive differences between the three initialization points. Hence, as stated previously in Section~\ref{sec:vim_strategy}, the VIM will experience fast synchronization for any reasonable guess of the initial rotor frequency prior to the grid connection.

\begin{figure}[!b]
    \centering
	\scalebox{1.125}{\includegraphics[]{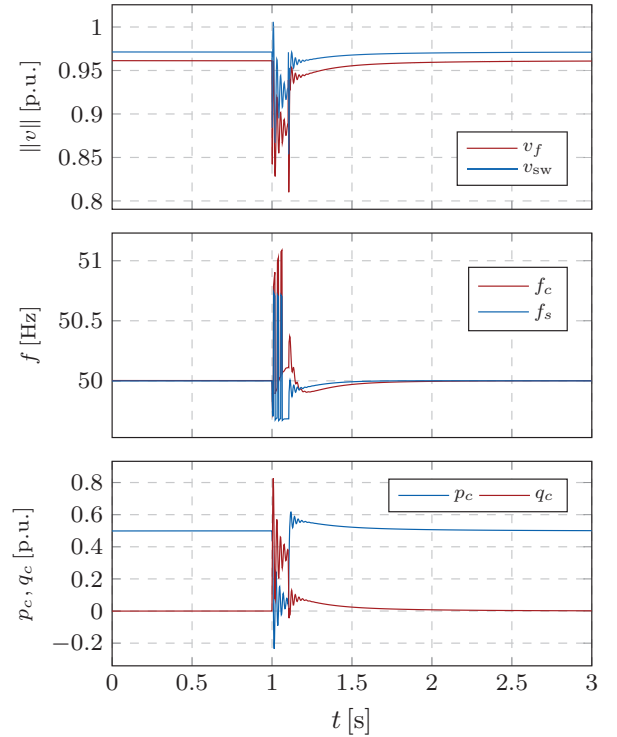}} 
	\caption{VIM operation during a short-circuit: (i) voltages; (ii) frequencies; (iii) active and reactive power.}
	\label{fig:short_circuit}
\end{figure}
\subsection{Resilience to Short Circuit Events}
Short circuit faults are the most common and severe events in power systems. Therefore, it is instrumental to study the resilience of VIM-synchronized grid-following units to these faults. To this end, the VSC is initialized to a steady-state operating point, with the voltage reference set at $V_c^\star = 1 \, \mathrm{p.u.}$, and the active and reactive power setpoints at $p_c^\star = 0.5 \, \mathrm{p.u.}$ and $q_c^\star = 0 \, \mathrm{p.u.}$, respectively. At $t=\SI{1}{\second}$, a three-phase short circuit is applied at the load bus (bus 3), and the fault is cleared after \SI{150}{\milli\second}. Responses of relevant VSC variables are presented in Fig.~\ref{fig:short_circuit}. As expected, the fault is followed by fast transients and a significant drop in voltages at the VSC terminal. During the event, the frequency of the VSC oscillates with a magnitude of \SI{1}{\hertz} without the loss of stability. After the fault clearing, the VSC manages to promptly re-synchronize to the grid, and the voltage and power variables continue to follow the prescribed setpoints. This study demonstrates the capability of a VIM-based VSC to remain stable and in synchronism during and after a short circuit event.

\subsection{Islanding and Frequency Disturbance Events}
Previous case studies focused on typical operational circumstances related to control of VSCs, such as grid-connection, setpoint changes, etc. The behavior of PLL-based grid-following units under these events is similar and well studied by the literature. Therefore, a detailed analysis of PLL-based units has been omitted. In this subsection, we compare the synchronization performance of grid-forming, and VIM and PLL-based grid-following VSCs in cases of two events: firstly, after the disconnection from the network, and secondly, during a frequency disturbance event. We would like to note that the same values have been used for all the control parameters common to all three unit types, such as droop gains, etc.
\begin{figure}[!b]
    \centering
	\scalebox{1.125}{\includegraphics[]{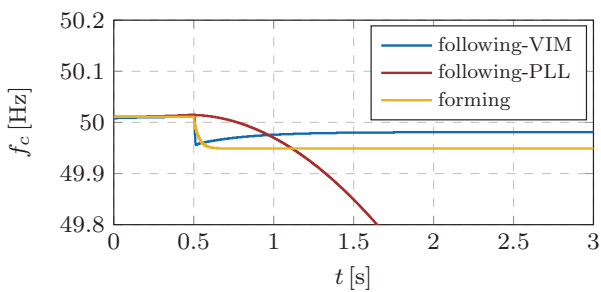}} 
	\caption{Behavior of different synchronization units after grid disconnection.}
	\label{3.1fig:disconnect}
\end{figure}

The islanding is simulated\footnote{We assume $t=0$ to be the time instance at which the initialization transients have decayed and the system is in synchronism.} at $t=\SI{0.5}{\second}$ and the inverter response is showcased in Fig.~\ref{3.1fig:disconnect}. The PLL-based unit loses synchronism immediately after the disconnection, with frequency plummeting below \SI{49}{\hertz} within \SI{3}{\second} of the disconnection. On the other hand, after some negligible initial transients, the VIM restores the converter to a new steady-state point and proceeds with normal operation, similarly to the grid-forming unit. Therefore, one of the main benefits of employing a VIM synchronization scheme lies in its standalone capability, i.e., the ability to operate even after being disconnected from the main grid if desirable; a characteristic of a physical induction machine. Such property is clearly not attainable by traditional grid-following VSCs employing a phase-locked loop.

\begin{figure}[!t]
    \centering
	\scalebox{1.125}{\includegraphics[]{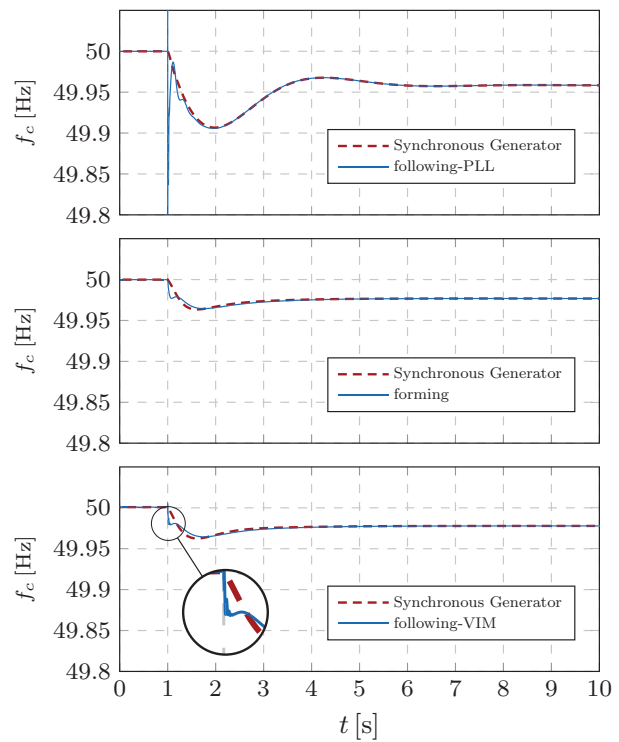}} 
	\caption{Behavior of unit frequencies after a load change event at bus 3: (a) interaction between SG and PLL-based grid-following unit; (b) interaction between SG and grid-forming VSC; (c) interaction between SG and VIM-based grid-following unit.}
	\label{fig:frequency_disturbance}
\end{figure}

To simulate a frequency disturbance event, the stiff grid equivalent at bus $2$ has been replaced by a $6^\text{th}$-order synchronous machine model from \cite{Kundur}. 
A low-inertia scenario is created by equating the installed capacities of both units. The load impedance is set to consume $1\,\mathrm{p.u.}$ at nominal voltage, and both VSC and SG are set to supply half of the load. A frequency disturbance event is simulated by applying a load impedance increase of $0.05\,\mathrm{p.u.}$ at $t=\SI{1}{\second}$, and frequency responses of SG, grid-forming, PLL- and VIM-based grid-following inverters have been showcased in Fig.~\ref{fig:frequency_disturbance}. After short initial transients, the PLL-based inverter manages to promptly synchronize to the varying frequency signal, while the VIM-based unit synchronizes slowly, after about \SI{1}{\second}. Nevertheless, the frequency response with the VIM-based VSC achieves a significant improvement in damping, similar to the response of the system with the grid-forming unit. Improvement in nadir is also observed, which indicates that using VIM might prevent underfrequency load shedding in cases of large system disturbances. 
In addition, the VIM-synchronization brings a reduction in the maximum instantaneous Rate-Of-Change-Of-Frequency (RoCoF) at the SG node. The maximum instantaneous RoCoF observed at the SG node in Fig.~\ref{fig:frequency_disturbance}-(a) is $\SI[per-mode=symbol]{-0.22}{\hertz\per\second}$ and in Fig.~\ref{fig:frequency_disturbance}-(c) is $\SI[per-mode=symbol]{-0.11}{\hertz\per\second}$. Therefore, apart from improving the frequency nadir, the VIM-based VSC reduces the maximum instantaneous RoCoF compared to a PLL-based unit. This result is significant as it is well-known that synchronous generators might experience mechanical stress when subjected to high RoCoF values~\cite{Kundur}.

\begin{figure}[!t]
    \centering
	\scalebox{1.125}{\includegraphics[]{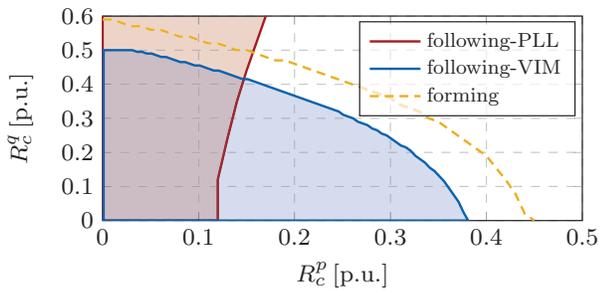}} 
	\caption{Stability maps of different converter modes in the $R_c^p-R_c^q$ plane. Shaded areas correspond to stable regions of PLL and VIM-based converters, whereas the region left from the yellow line is stable for forming VSCs. }
	\label{3.1fig:stabMapDroop}
\end{figure}
\section{Stability Analysis} \label{3.1sec:stab_analysis}

Having validated the theoretical concept of VIM through EMT simulations, we dedicate this section to small-signal analysis of the DAE model presented in Section~\ref{sec:modeling}. Moreover, we consider three different converter operation modes: (i) a grid-following VSC with a PLL; (ii) a grid-following VSC with a VIM; and (iii) a grid-forming VSC from Sec.~\ref{subsec:sys_level_form}.

\subsection{Droop Gain Stability Map}
Some very insightful observations can be made by studying the stability maps in the $R_c^p-R_c^q$ plane of a single inverter connected to a stiff grid. One such map is illustrated in Fig.~\ref{3.1fig:stabMapDroop} for a wide range of active and reactive power droop gains. Clearly, the two grid-following controllers have significantly different stability regions. However, for a reasonable tuning range considered in practice (i.e., $R_c^p<0.1\,\mathrm{p.u.}$ and $R_c^q<0.05\,\mathrm{p.u.}$), the two regions are identical. Nevertheless, a very interesting observation can be made by comparing the aforementioned stability maps to the corresponding map of a grid-forming unit. Indeed, the stable region of a grid-forming VSC closely resembles the one of a VIM-based inverter for the whole parameter range, suggesting that replacing a PLL with a VIM-based synchronization device might provide some ``forming-like'' properties to the grid-following VSCs. A notion that could be substantiated by the fact that the VIM has standalone capabilities.

\begin{figure}[!b]
    \centering
	\scalebox{1.125}{\includegraphics[]{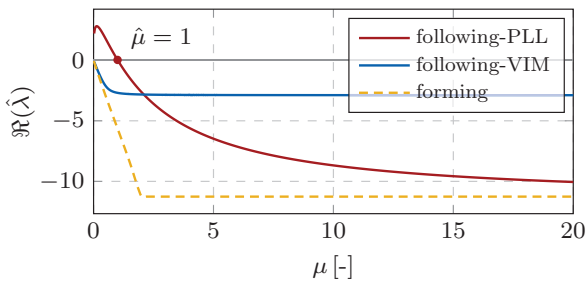}} 
	\caption{SCR influence on system stability for different converter modes.}
	\label{fig:SCR}
\end{figure}

\subsection{Impact of the Grid SCR and Permissible Penetration Level}
\begin{figure}[!b]
    \centering
	\scalebox{1.125}{\includegraphics[]{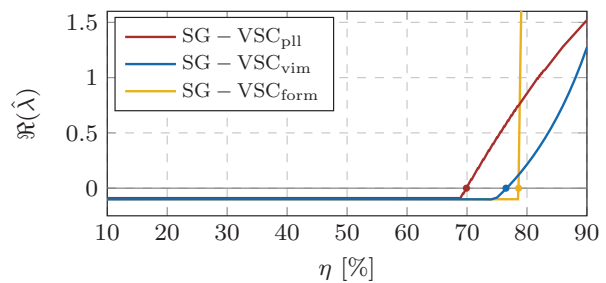}} 
	\caption{Impact of penetration of inverter-based generation on system stability for different converter operation modes.}
	\label{fig:eig2bus}
\end{figure}

We continue this line of investigation by comparing the stability margins of all three converter modes depending on the strength (i.e., the short-circuit ratio $\mu\in\R$) of the grid at the connection terminal. As can be seen from Fig.~\ref{fig:SCR}, the critical SCR for a PLL-based inverter is $\hat{\mu}=1$, whereas the grid-forming VSC does not impose any requirements on the minimum grid strength. In that sense, the VIM-controlled inverter again resembles the grid-forming one, since it can withstand any SCR level. Moreover a similar analysis is done with respect to the maximum permissible penetration of inverter-interfaced generation. The movement of the most critical eigenvalue, i.e., the evolution of its real part with the increase in VSC installation level $\eta\in[0,100]\,\%$, is depicted in Fig.~\ref{fig:eig2bus} for the 3-bus test case, with the stiff grid equivalent at bus 2 replaced by an SG. While the maximum permissible installed capacity of PLL-based units is slightly below \SI{70}{\percent}, this level can be increased by $\approx\SI{7}{\percent}$ by substituting the PLL with the VIM, therefore almost reaching the maximum permissible penetration of grid-forming VSCs of $\eta_\mathrm{max}=\SI{78.5}{\percent}$. Note that the eigenvalue movement for scenarios $\mathrm{SG}-\mathrm{VSC}_\mathrm{pll}$ and $\mathrm{SG}-\mathrm{VSC}_\mathrm{form}$ is adopted from \cite{UrosStability2021}. This indicates that, despite not being able to provide black start and independently generate stable frequency reference, the VIM-based grid-following converters clearly share conceptual similarities with the grid-forming mode of operation. Furthermore, it should be pointed out that the instabilities arising at the maximum converter installation levels obtained from Fig.~\ref{fig:eig2bus} are solely associated with device-level control, with the highest state participation given in Table~\ref{tab:partfac}. Hence, the instability is related to the inner voltage and current control and is not affected by the selection of the synchronization unit.
\begin{table}[!t]
\renewcommand{\arraystretch}{2}
\caption{State participation $[\%]$ in the unstable modes.}
\label{tab:partfac}
\noindent
\centering
    \begin{minipage}{\linewidth} %Use the minipage environment to footnote tables
    \renewcommand\footnoterule{\vspace*{-5pt}} %to remove the horizontal rule above the table footnote
    \begin{center}
    \scalebox{0.9}{%
        \begin{tabular}{ c || c | c | c | c | c | c | c | c | c }
            \toprule
            \textbf{State} & $i_f^d$ & $i_f^q$ & $v_f^d$ & $v_f^q$ & $\xi^d$ & $\xi^q$ & $i_g^d$ & $i_g^q$ & other\\ 
            \cline{1-10}
            \textbf{PF} & $30.2$ & $30.5$ & $8.1$ & $8.4$ & $6.2$ & $6.8$ & $4.5$ & $4.1$ & $1.2$ \\              
            \arrayrulecolor{black}\bottomrule
        \end{tabular}
    }
        \end{center}
    \end{minipage}
\end{table}

\subsection{Recommendations for VIM Tuning}
\begin{figure}[!t]
    \centering
	\scalebox{1.1}{\includegraphics[]{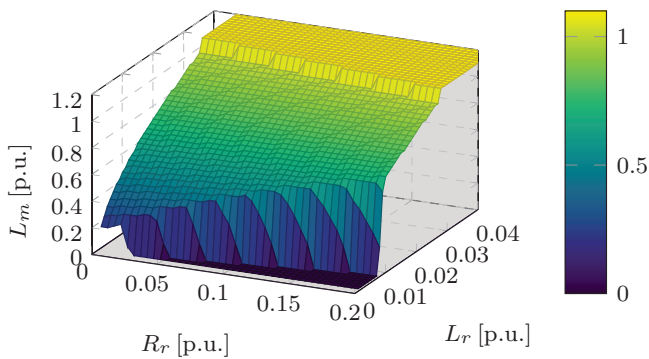}} 
	\caption{Stability surface of a VIM-based converter in the $R_r-L_r-L_m$ space. The shaded area below the surface indicates a stable operating region.}
	\label{3.1fig:stabSurf}
\end{figure}

Finally, we briefly address the topic of parameter tuning pertaining to the proposed VIM design. As previously pointed out in Section~\ref{sec:vim_strategy}, the tuning of a VIM can be based on the parameters of a physical induction machine. This particularly applies to the selection of resistance and inductance values involved in the PD controller $K_\nu(s)$ in \eqref{eq:Wslip} and the transfer function $K_e(s)$ in \eqref{eq:Ke}. However, emulating the existing machine has its drawbacks, as it does not guarantee an optimal control performance. This was already discussed in Section~\ref{3.1sec:sim_results}, as the the derivative gain $K_\nu^D$ had to be re-tuned in order to achieve a satisfactory response during transients. Moreover, some induction generators might simply result in an unstable system. One such example is provided in Fig.~\ref{3.1fig:stabSurf} with the stability surface of a VIM-based converter illustrated in the $R_r-L_r-L_m$ space. It reflects the possible obstacles one could face when applying heuristic methods for VIM tuning, as the surface in Fig.~\ref{3.1fig:stabSurf} has distinctively nonlinear segments. Nevertheless, since VIM is an emulation of a physical machine, the tuning process does not have to incorporate exact physical parameters, but can rather use them as an initial starting point when designing the controller. Therefore, the multi-dimensional stability mapping, such as the one illustrated in Fig.~\ref{3.1fig:stabSurf}, can be of great importance when optimizing the VIM performance through parameter tuning.

\begin{figure}[!b]
    \centering
    \includegraphics[scale=0.75]{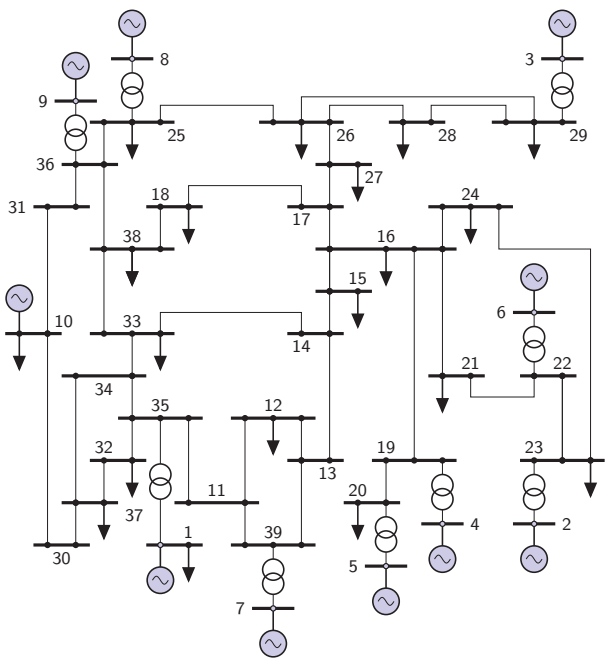}
    \caption{Single line diagram of the IEEE 39-bus test system.}
    \label{fig:39busIEEE}
\end{figure}
\subsection{Stability Analysis of 39-bus IEEE Test System}
In order to study stability characteristics of large-scale low-inertia systems with deployed VIM-based VSCs, we consider the 39-bus IEEE system presented in Fig.~\ref{fig:39busIEEE}. This is a 10-machine representation of the New England power system, with the generator at node 10 representing the aggregation of a large number of generators. For more information on the network, and relevant load and generation parameters, the reader is referred to \cite{Athay1979,39busBook}.

The stability analysis is carried out by assesssing the small-signal stability of different configuration scenarios in which conventional SG-based units are replaced by grid-following VSCs synchronized through PLL or VIM. The list of all considered scenarios is given in Table~\ref{tab:stab_scenarios}. The letter ``S'' is used to designate a synchronous generator, whereas the letter ``f'' denotes a grid-following converter. The first column is the scenario identifier (ID), followed by ten columns showcasing generation types at the respective nodes. Finally, stability of the corresponding synchronization method is indicated in the last two columns.

The considered nine scenarios were constructed by gradually replacing the synchronous generation with grid-following VSCs to assess their maximum admissible penetration. Trivial generation portfolios based entirely on synchronous generation (stable) or entirely on grid-following VSCs (unstable) are not presented. Using the results from the table, we can conclude that for moderate penetration levels of grid-following VSCs, both synchronization methods can achieve small-signal stability. However, unlike PLL, the VIM synchronization method allows generation portfolios strongly dominated by grid-following VSCs. In the observed case study, penetration levels up to $70\%$ are admissible when using PLL synchronization, whereas VIM synchronization allows $85\%$ penetration. For brevity, grid-forming VSCs were not considered in this case study. However, the work in \cite{UrosStability2021} demonstrated that grid-forming units could achieve over $90\%$ penetration in large-scale power systems. Note that the percentages slightly change by considering different permutations of the generation portfolios, different load models, etc. Nonetheless, the main conclusion that the VIM-based synchronization enables higher penetration remains.
\begin{table}[!t]
\def\arraystretch{1.15}
\caption{Configuration of generation portfolio scenarios and stability analysis of the 39-bus IEEE test system.}
    \centering
\resizebox{0.95\linewidth}{!}{%
            \begin{tabular}{ c || c | c | c | c | c | c | c | c | c | c || c | c }
            \toprule
            \multirow{2}{*}{\textbf{ID}} & \multicolumn{9}{c}{\textbf{Generator Type at Node}} & & \multicolumn{2}{c}{\textbf{Sync.}}\\
            \cline{2-13}
             &1&2&3&4&5&6&7&8&9&10& PLL& VIM \\
            \hline
            1& f& S& S& S& S& S& S& S& S& S&\cellcolor{myGreen!25} stab &\cellcolor{myGreen!25} stab\\
            2& f& f& S& S& S& S& S& S& S& S&\cellcolor{myGreen!25} stab &\cellcolor{myGreen!25} stab\\
            3& f& f& f& S& S& S& S& S& S& S&\cellcolor{myGreen!25} stab &\cellcolor{myGreen!25} stab\\
            4& f& f& f& f& S& S& S& S& S& S&\cellcolor{myGreen!25} stab &\cellcolor{myGreen!25} stab\\
            5& f& f& f& f& f& S& S& S& S& S&\cellcolor{myGreen!25} stab &\cellcolor{myGreen!25} stab\\
            6& f& f& f& f& f& f& S& S& S& S&\cellcolor{myRed!25} unstab &\cellcolor{myGreen!25} stab\\
            7& f& f& f& f& f& f& f& S& S& S&\cellcolor{myRed!25} unstab &\cellcolor{myGreen!25} stab\\
            8& f& f& f& f& f& f& f& f& S& S&\cellcolor{myRed!25} unstab &\cellcolor{myRed!25} unstab\\
            9& f& f& f& f& f& f& f& f& f& S&\cellcolor{myRed!25} unstab &\cellcolor{myRed!25} unstab\\
            \arrayrulecolor{black!30}\cline{2-13}
            \arrayrulecolor{black}\bottomrule
        \end{tabular}
        }
    \vspace{0.2cm}
    \label{tab:stab_scenarios}
\end{table}

\section{Conclusion} \label{sec:5}

This paper proposes a novel control strategy for the synchronization of grid-following VSCs based on the emulation of induction generator principles. The proposed formulation allows seamless integration of the proposed synchronization method in the grid-following control structure by simply replacing the PLL unit. The EMT simulations showcase smooth start-up and synchronization to the grid and accurate computation of grid frequency, independent of the initial rotor speed input. Furthermore, the proposed synchronization device does not hinder the performance of other converter controls and preserves accurate voltage and power setpoint tracking. It is shown that in contrast to PLL, VIM-synchronized converters can operate in standalone mode after an islanding event and are resilient to short circuit events. Our analysis further reveals that the VIM synchronization may aid frequency containment in low-inertia systems and lead to an improved damping of the frequency response. Finally, small-signal stability results show that a VIM-based grid-following converter resembles a grid-forming unit in certain operational aspects, allowing for higher penetration of VIM-controlled converters than the PLL-based ones. In other words, simply replacing the commonly used PLL synchronization with VIM may lead to substantial improvement in the system stability margins.

On the other hand, several potential weaknesses of the proposed synchronization method have been identified. Firstly, the structure of the VIM is complex and relies on more parameters than PLL. Secondly, while PLL relies only on the terminal voltage measurement, VIM in addition requires the three-phase current measurement at the VSC output, therefore being more prone to measurement delays and errors. Thirdly, an integral part of the VIM is a PD controller that might be prone to overshoots and even instability if not properly tuned. Furthermore, grid synchronization has generally shown to be slower when using VIM compared to PLL. Finally, in contrast to grid-forming converters, neither PLL nor VIM synchronization possesses the black-start capability in the grid-following mode.

An important avenue for future work is the analysis of the impact of measurement delays and errors on the operation of VIM-based grid-following VSCs. Furthermore, synchronization performance under unbalanced and distorted grid currents needs to be assessed. To ensure the practicality of the proposed approach, the behavior of VIM under phase jumps and unbalanced grid conditions should be evaluated. Finally, our future work will attempt to perform extensive hardware-in-the-loop tests to validate the proposed approach.
% \vspace{-0.25cm}
% Appendix
\appendices
\section{Derivation of Saturated Frequency Slip} \label{sec:app_3.1}

Let us first recall the expressions for the minimum and maximum of two variables from \eqref{eq:minmax}.
The goal is to impose the lower and upper saturation limits $(\ushort{\omega}_\nu,\widebar{\omega}_\nu)$ on an unsaturated frequency slip signal $\tilde{\omega}_\nu$, i.e., to obtain a new signal $\omega_\nu$ such that $\omega_\nu\in\left[\ushort{\omega}_\nu,\widebar{\omega}_\nu\right]$. This is equivalent to finding the maximum of the underlying signal and its lower bound (let us denote it by $\hat{\omega}_\nu$), and subsequently finding the minimum of that signal and the upper bound. In other words, 
\begin{subequations}
\label{3.1eq:minmaxW}
\begin{align}
    \hat{\omega}_\nu = \max\left\{\tilde{\omega}_\nu,\ushort{\omega}_\nu\right\}, \label{3.1eq:minmaxW1} \\
    \omega_\nu = \min\left\{\hat{\omega}_\nu,\widebar{\omega}_\nu\right\}. \label{3.1eq:minmaxW2}
\end{align}
\end{subequations}
We can rewrite \eqref{3.1eq:minmaxW} as:
\begin{subequations}
\begin{align}
    \hat{\omega}_\nu = \frac{1}{2} \left( \tilde{\omega}_\nu + \ushort{\omega}_\nu +\left| \ushort{\omega}_\nu - \tilde{\omega}_\nu \right| \right), \label{eq:slip_sat_new1}\\
    \omega_\nu = \frac{1}{2} \left( \hat{\omega}_\nu + \widebar{\omega}_\nu - \left| \widebar{\omega}_\nu - \hat{\omega}_\nu \right|\right), \label{eq:slip_sat_new2}
\end{align}
\end{subequations}
and subsequently substitute \eqref{eq:slip_sat_new2} into \eqref{eq:slip_sat_new1}, resulting in the expression given by \eqref{eq:slip_sat}.

% References section
\bibliographystyle{IEEEtran}
\bibliography{bibliography}

\end{document}